\documentclass[preprint,aps,amsmath,showpacs,showkeys,nofootinbib, 
superscriptaddress]{revtex4}
\usepackage{epsfig}
\usepackage{graphicx}
\usepackage{multirow}
\usepackage{tablefootnote}

\begin{document} 

\title{\textbf{\bf The ${}_{\Lambda\Lambda}^{\,\,\,\,4}n$ system}}
\author{H.~Garcilazo} 
\email{humberto@esfm.ipn.mx} 
\affiliation{Escuela Superior de F\' \i sica y Matem\'aticas, \\ 
Instituto Polit\'ecnico Nacional, Edificio 9, 
07738 M\'exico D.F., Mexico} 
\author{A.~Valcarce} 
\email{valcarce@usal.es} 
\affiliation{Departamento de F\'\i sica Fundamental and IUFFyM,\\ 
Universidad de Salamanca, E-37008 Salamanca, Spain}
\author{J.~Vijande}
\email{javier.vijande@uv.es}
\affiliation{Unidad Mixta de Investigaci\'on en Radiof{\'\i}sica e Instrumentaci\'on Nuclear en Medicina (IRIMED),
Instituto de Investigaci\'on Sanitaria La Fe (IIS-La Fe)-Universitat de Valencia (UV) and IFIC (UV-CSIC), Valencia, Spain}
\date{\today} 
\begin{abstract}
Using local central Yukawa-type Malfliet-Tjon interactions reproducing 
the low-energy parameters and phase shifs of the $nn$ system and the 
latest updates of the $n\Lambda$ and $\Lambda\Lambda$ Nijmegen ESC08c 
potentials we study the possible existence of a 
${}_{\Lambda\Lambda}^{\,\,\,\,4}n$ bound state.
Our results indicate that the ${}_{\Lambda\Lambda}^{\,\,\,\,4}n$
is unbound, being just above threshold. We discuss the role 
played by the $^1S_0$ $nn$ repulsive term of the Yukawa-type Malfliet-Tjon interaction. 
\end{abstract}
\pacs{21.45.-v,25.10.+s,11.80.Jy}
\keywords{baryon-baryon interactions, few-body systems} 
\maketitle 

\section{Introduction} 
It is well-established the non-existence of two-body bound states made
of neutrons and/or $\Lambda$'s, the lightest hyperon. The situation is 
much more cumbersome for three-, four- and in general few-body systems
made of nucleons and hyperons~\cite{Ric15}. For example,
it has been proposed that dineutrons could become bound in the presence of
additional nucleons~\cite{Mig73}. This is the mechanism responsible for
the properties of some bound nuclei that have a neutron excess, like $^{11}$Li,
where a pair of external neutrons form a remote halo around the core of $^9$Li~\cite{Ber16}.
Such possibility has been recently drawn in a lighter system 
by the experimental HypHI Collaboration~\cite{Rap13}, suggesting 
the existence of a neutral bound state of two neutrons and 
a $\Lambda$ hyperon, $^3_\Lambda n$. They analyze the
experimental data obtained from the reaction $^6$Li +$^{12}$C at 2A GeV
to study the invariant mass distribution of $d+\pi^-$ and $t+\pi^-$.
The signal observed in the invariant mass distributions of $d+\pi^-$ and $t+\pi^-$ 
final states was attributed to a strangeness-changing weak process corresponding
to the two- and three-body decays of an unknown bound state of two neutrons associated
with a $\Lambda$, $^3_\Lambda n$, via $^3_\Lambda n \to t + \pi^-$ and
$^3_\Lambda n \to t^* + \pi^- \to d + n + \pi^-$.
This is an intriguing conclusion since one would naively expect the $nn\Lambda$ system
to be unbound. In the $nn\Lambda$ system the two nucleons interact in the $^1S_0$
partial wave while in the $np\Lambda$ system they interact in the $^3S_1$ partial wave.
Thus, since the nucleon-nucleon ($NN$) interaction in the $^1S_0$ channel is weaker than the $^3S_1$ channel,
and the $np\Lambda$ system is bound by only 0.13 MeV, one may have anticipated that the
$nn\Lambda$ system should be unbound. The absence of binding of the $nn\Lambda$ system was first 
demonstrated by Dalitz and Downs~\cite{Dal58} using a variational approach, and
later from the solution of
the Faddeev equations with separable interactions \cite{Gar87}.
The theoretical debate on the possible existence of a neutral bound state
of two neutrons and a $\Lambda$ hyperon, $^3_\Lambda n$, is still
open and has lately deserved an important theoretical 
effort~\cite{Gar14,Hiy14,Gal14,And15,Afn15,Ric15}.

In the four-body case, the analysis of the missing-mass spectrum in the double-charge-exchange
reaction $^4$He($^8$He,$^8$Be) at 186 MeV/u has unveiled the possible existence
of a tetraneutron resonance $0.83 \pm 0.65 \, {\rm (stat)} \, \pm 1.25 \, {\rm (syst)}$ MeV above the
threshold of four-neutron decay with a significance level of 4.9$\,\sigma$~\cite{Kis16}.
In 2002 one collaboration claimed to have found a bound tetraneutron
in a $^{14}$Be breakup reaction~\cite{Mar02}. This result remains unconfirmed,
and theorists quickly showed that based on the best knowledge of the $NN$
interaction the existence of a bound tetraneutron was nearly impossible,
although they could not rule out the existence of a short-lived resonant
state on the basis of a dineutron-dineutron structure~\cite{Ber03, Pie03,Tim03,Laz05}.
The stability of a tetraneutron state cannot be established
even with potentials made artificially deeper to produce a dineutron
bound state (the dineutron is a virtual state 66 keV above the two-neutron
threshold), due to the Pauli principle which forbids two identical
fermions from occupying the same quantum state. For four-neutrons
only one pair can be in the lowest-energy state, forcing the
second pair into a state of higher energy, thereby making the tetraneutron
unstable. Thus, one could think of the stability of a modified tetraneutron
with Bose statistics, where a pair of neutrons is replaced by a pair
of neutral light baryons enforcing in this way antisymmetrization with
all particles in the lowest-energy state. This is the case
of the ${}_{\Lambda\Lambda}^{\,\,\,\,4}n$ = $(n,n,\Lambda,\Lambda)$
recently discussed in Ref.~\cite{Ric15} and suggested as a possible Borromean
state.

The relevance of the addition of further baryons on an almost bound two-body system
has also been discussed recently by some
of us looking for stable bound states of $N$'s and $\Xi$'s. In Ref.~\cite{Gar15} we
pointed out that when a two-baryon interaction is attractive, if the system is merged with nuclear matter
and the Pauli principle does not impose severe restrictions, the attraction may be reinforced.
Simple examples of the effect of a third or a fourth baryon in two-baryon
systems could be given. The deuteron, $(I)J^P=(0)1^+$, is bound by $2.225$ MeV, while the triton,
$(I)J^P=(1/2)1/2^+$, is bound by $8.480$ MeV, and the $\alpha$ particle, $(I)J^P=(0)0^+$,
is bound by $28.295$ MeV. The binding per nucleon $B/A$ increases from $1:3:7$.
A similar argument could be employed for strangeness $-1$ systems. Whereas there is 
no evidence for dibaryon states, the hypertriton $^3_\Lambda$H, $(I)J^P=(0)1/2^+$, is bound with a separation
energy of $130 \pm 50$ keV, and the $^4_\Lambda$H, $(I)J^P=(0)0^+$, is bound
with a separation energy of $2.12 \pm 0.01 \, {\rm (stat)} \, \pm 0.09 \, {\rm (syst)}$ MeV~\cite{Ess15}. 
This cooperative effect of the attraction in 
the two-body subsystems when merged in few-baryon states was also made evident
in the prediction of a $\Sigma NN$ quasibound state in the $(I)J^P = (1)1/2^+$ 
channel very near threshold~\cite{Gar07}.
Such $\Sigma NN$ quasibound state has been recently suggested
in $^3\rm{He}(K^-,\pi^\mp)$ reactions at 600 MeV/c~\cite{Har14}.

Thus, if a second $\Lambda$ would be added to the uncertain $nn\Lambda$ state, 
the weakly attractive $\Lambda\Lambda$ interaction~\cite{Tak01} and the reinforcement 
of the $N\Lambda$ potential without paying a price for antisymmetry requirements, 
may give rise to a stable bound state. This would be our goal in this paper, 
to address the study of the ${}_{\Lambda\Lambda}^{\,\,\,\,4}n$ state making use of potentials
compatible with the low-energy data and phase-shifts of the $nn$, $n\Lambda$, and $\Lambda\Lambda$
systems. A first examination of this problem has been presented in Ref.~\cite{Ric15}
based on potentials with a single Yukawa attractive term or a Morse parametrization. 

\section{Two-body interactions} 

For the identical pairs,
$nn$ and $\Lambda\Lambda$, the S wave interaction is in the $^1S_0$
channel due to the Pauli principle, while for the $N\Lambda$ pair
both $^1S_0$ and $^3S_1$ channels contribute. As it is
well-known, the $NN$ $^1S_0$ channel is almost bound, the virtual state 
lying slightly below the $nn$ threshold in the
unphysical sheet. In the case 
of the $NN$ $^1S_0$ channel we use
the Malfliet-Tjon I model~\cite{Mal69} with the parameters given in Ref.~\cite{Gib90}.
For the two-body interactions containing $\Lambda$'s, $N\Lambda$ and $\Lambda\Lambda$, we use the most recent update
on the ESC08c Nijmegen potentials~\cite{Nae15,Nag15,Rij16}.
Regarding the two-body interactions containing a 
single $\Lambda$, they are constrained by a simultaneous fit
to the combined $NN$ and $YN$ scattering data, supplied with 
constraints on the $YN$ and $YY$ interaction originating from 
the G-matrix information on hypernuclei~\cite{Nae15}.
The $\Lambda\Lambda$ strangeness $-2$ interaction
is mainly determined by the $NN$ and $YN$ data, and SU(3)
symmetry~\cite{Nag15,Rij16}. It gives account of the pivotal
results of strangeness $-2$ physics, the NAGARA~\cite{Tak01} 
and the KISO~\cite{Naa15} events. Although other double-$\Lambda$
hypernuclei events, like the DEMACHIYANAGI and HIDA events~\cite{Nak10},
are not explicitly taken into account, the G-matrix nuclear matter study
of $\Xi^-$ capture both in $^{12}$C and $^{14}$N (see section
VII of Ref.~\cite{Nag15}), concludes that the $\Xi N$ attraction
in the ESC08c potential is consistent with the $\Xi$-nucleus binding 
energies given by the emulsion data of the twin $\Lambda$-hypernuclei.

We have constructed the two-body amplitudes for all subsystems entering the four-body 
problem studied by solving the Lippmann--Schwinger
equation of each $(i,j)$ channel,
\begin{equation}
t^{ij}(p,p';e)= V^{ij}(p,p')+\int_0^\infty {p^{\prime\prime}}^2
dp^{\prime\prime} V^{ij}(p,p^{\prime\prime})
\frac{1}{e-{p^{\prime\prime}}^2/2\mu} t^{ij}(p^{\prime\prime},p';e) \, ,
\label{eq19} 
\end{equation}
where 
\begin{equation}
V^{ij}(p,p')=\frac{2}{\pi}\int_0^\infty r^2dr\; j_0(pr)V^{ij}(r)j_0(p'r) \, ,
\label{eq20} 
\end{equation}
and the two-body potentials consist of an attractive and a repulsive
Yukawa term, i.e.,
\begin{equation}
V^{ij}(r)=-A\frac{e^{-\mu_Ar}}{r}+B\frac{e^{-\mu_Br}}{r} \, .
\label{eq21} 
\end{equation}
The parameters of the $\Lambda N$ and $\Lambda \Lambda$ 
channels were obtained by fitting the low-energy data and the phase-shifts of each channel as given 
in the most recent update of the strangeness $-1$~\cite{Nae15} and $-2$~\cite{Nag15} ESC08c Nijmegen 
potential. The low-energy data and the parameters of these models, together
with those of the $NN$ interaction from Ref.~\cite{Gib90}, are given in Table~\ref{t1}.
It is worth to note that the scattering length and effective range of the most recent
updates of the $\Lambda\Lambda$ interaction derived from chiral effective field
theories are very much like those of the ESC08c Nijmegen potential (see Table 2 of
Ref.~\cite{Hai16}) unlike the earlier version 
used in Ref.~\cite{Ric15} (see Table 4 of Ref.~\cite{Pol07})
reporting remarkably small effective ranges.
\begin{table}[t]
\caption{Low-energy parameters and parameters of the local central Yukawa-type potentials 
given by Eq.~(\ref{eq21}) for the $NN$~\cite{Gib90}, $\Lambda N$~\cite{Nae15}, and $\Lambda \Lambda$~\cite{Nag15} systems
contributing to the $(I)J^P=(1)0^+$ ${}_{\Lambda\Lambda}^{\,\,\,\,4}n$
state. See text for details.} 
\begin{ruledtabular} 
\begin{tabular}{ccccccccc} 
& $(i,j)$ & $a({\rm fm})$ & $r_0({\rm fm})$ & $A$(MeV fm) & 
$\mu_A({\rm fm}^{-1}$) 
& $B$(MeV fm) & $\mu_B({\rm fm}^{-1})$  & \\
\hline
\multirow{1}{*}{$NN$} & $(1,0)$  & $-23.56$ & $2.88$ & $513.968$  & $1.55$  & $1438.72$ & $3.11$ & \\
\multirow{2}{*}{$\Lambda N$}& $(1/2,0)$ & $-2.62$  & $3.17$  &  $416$  & $1.77$  & $1098$ & $3.33$ & \\ 
& $(1/2,1)$ & $-1.72$  & $3.50$  &  $339$  & $1.87$  & $968$ & $3.73$ & \\ \hline
\multirow{1}{*}{$\Lambda \Lambda$} & $(0,0)$ & $-0.853$& $5.126$ &  $121$  & $1.74$  & $926$ & $6.04$ & \\ 
\end{tabular}
\end{ruledtabular}
\label{t1} 
\end{table}

If it is assumed that only singlet and triplet S wave contribute in the
two-particle channel, the parametrization of the $NN$ interaction used 
in this work, set III for the triplet partial wave and set I for the 
singlet partial wave, gives a triton binding energy of 8.3 MeV~\cite{Mal69}. 
The effect of the repulsive core on the singlet two-body channel is crucial 
to get this result, while the repulsion on the triplet two-body channel has 
almost no effect on the binding. In fact, if the repulsive core in the singlet 
partial wave is not considered the triton gains around 2 MeV of binding (see Table II of 
Ref.~\cite{Mal70}). Based on predictions for separable potentials, 
in Ref.~\cite{Mal69} it is suggested that the inclusion of the tensor force 
in the triplet interaction changes the binding energy by 0.3 MeV. Indeed, 
this is the result obtained in Ref.~\cite{Fuj02}, where as can be seen in 
Table III a five channel calculation (S and D partial waves) 
differs from a two channel calculation (only S partial waves) about 0.3 MeV.
The influence of local tensor forces in Malfliet-Tjon Yukawa type
interactions has also been studied in Ref.~\cite{Maf69}, showing that the inclusion of
tensor forces reduces the binding energy of the three-body problem by 1 to
1.5 MeV, depending on the D wave percentage. Thus, the local Yukawa-type potentials
with tensor interaction would lack binding in the three-body problem
at difference of separable potentials that would drive to overbinding~\cite{Phi68}.
Note that in the ${}_{\Lambda\Lambda}^{\,\,\,\,4}n$ 
the $NN$ $^3S_1$ partial wave does not contribute, thus although this system is free
of any uncertainty related to the triplet partial wave, the repulsive core on 
the singlet $NN$ channel might play some role.

\section{The four-body problem} 
The four-body problem has been addressed by means of a generalized variational
method. The nonrelativistic hamiltonian will be given by,
\begin{equation}
H=\sum_{i=1}^4\frac{\vec p_{i}^{\,2}}{2m_{i}}+\sum_{i<j=1}^4V(\vec r_{ij}) \, ,
\label{ham}
\end{equation}
where the potentials $V(\vec r_{ij})$ have been discussed in the previous section.
For each channel $s$, the variational wave function will be the tensor product of
a spin ($\left|S_{s_1}\right>$), isospin ($\left|I_{s_2}\right>$), and radial
($\left|R_{s_3}\right>$) component,
\begin{equation}
\label{efr}
\left| \phi _{s}\right>=\left|S_{s_1}\right>\otimes\left|I_{s_2}\right>\otimes\left|R_{s_3}\right> \, ,
\end{equation}
where $s\equiv\{s_1,s_2,s_3\}$. Once the spin and isospin parts are integrated out, 
the coefficients of the radial wave function are
obtained by solving the system of linear equations,
\begin{equation}
\label{funci1g}
\sum_{s'\,s} \sum_{i} \beta_{s_3}^{(i)} 
\, [\langle R_{s_3'}^{(j)}|\,H\,|R_{s_3}^{(i)}
\rangle - E\,\langle
R_{s_3'}^{(j)}|R_{s_3}^{(i)}\rangle \delta_{s,s'} ] = 0 
\qquad \qquad \forall \, j\, ,
\end{equation}
where the eigenvalues are obtained by a minimization procedure.

For the description of the four-body wave function 
we consider the Jacobi coordinates:
\begin{eqnarray}
\label{coo}
& &\vec{r}_{NN}=\vec{x} =\vec{r}_{1}-\vec{r}_{2} \, , \nonumber \\
& & \vec{r}_{\Lambda\Lambda}=\vec{y} =\vec{r}_{3}-\vec{r}_{4} \, , \nonumber \\
& & \vec{r}_{NN-\Lambda\Lambda} = \vec{z} =\frac{1}{2} \left( \vec{r}_{1} + \vec{r}_{2} \right) -\frac{1}{2} 
\left( \vec{r}_{3}+\vec{r}_{4} \right) \, , \\
& &\vec{R}_{\rm CM} = \vec{R} =\frac{\sum m_{i}\vec{r}_{i}}{\sum m_{i}}\nonumber \, .
\end{eqnarray}
The total wave function should have well-defined permutation properties under
the exchange of identical particles. 
The most general S wave radial wave function may 
depend on the six scalar quantities that can be constructed with the Jacobi coordinates of the system, they are: 
$\vec x^{\,2}$, $\vec y^{\,2}$, $\vec z^{\,2}$, $\vec{x}\cdot\vec{y}$, $\vec{x}\cdot\vec{z}$ and $\vec{y}\cdot\vec{z}$. 
We define the variational spatial wave function as a linear combination of {\em generalized Gaussians},
\begin{equation}
\left|R_{s_3}\right>=\sum_{i=1}^{n} \beta_{s_3}^{(i)} R_{s_3}^i(\vec x,\vec y,\vec z)=\sum_{i=1}^{n} \beta_{s_3}^{(i)} R_{s_3}^i \, ,
\label{wave}
\end{equation}
where $n$ is the number of Gaussians used for each spin-isospin component. 
$R_{s_3}^i$ depends on six variational parameters:
$a^i_s$, $b^i_s$, $c^i_s$, $d^i_s$, $e^i_s$, and $f^i_s$, one for each scalar quantity. 
Therefore, the four-body system will depend on $6\times n\times n_s$ variational parameters, 
where $n_s$ is the number of different channels allowed by the Pauli principle.
Eq.~(\ref{wave}) should have well defined permutation symmetry under the exchange of both
$N$'s and $\Lambda$'s,
\begin{eqnarray}
\label{parx}
P_{12}(\vec x	\rightarrow -\vec x)R^i_{s_3}&=&P_xR^i_{s_3}\\ \nonumber
P_{34}(\vec y	\rightarrow -\vec y)R^i_{s_3}&=&P_yR^i_{s_3},
\end{eqnarray}
where $P_x$ and $P_y$ are $-1$ for antisymmetric states, $(A)$, and $+1$ for symmetric ones, $(S)$.
 
If we now define the function,
\begin{equation}
\label{red1}
g(s_1,s_2,s_3)={\rm Exp}\left(-a^i_s\vec x^{\,2}-b^i_s\vec y^{\,2}-c^i_s\vec z^{\,2}
-s_1d^i_s\vec x\cdot\vec y-s_2e^i_s\vec x\cdot\vec z-s_3f^i_s\vec y\cdot\vec z\right),
\end{equation}
and the vectors,
\begin{equation}
\vec{G}_s^i=\left(\begin{array}{l} g(+,+,+)\\g(-,+,-)\\g(-,-,+)\\g(+,-,-)\end{array}\right)\, ,
\end{equation}
and
\begin{eqnarray}
\label{red2}
\vec{\alpha}_{SS}&=&(+,+,+,+)\\ \nonumber
\vec{\alpha}_{SA}&=&(+,-,+,-)\\ \nonumber
\vec{\alpha}_{AS}&=&(+,+,-,-)\\ \nonumber
\vec{\alpha}_{AA}&=&(+,-,-,+),
\end{eqnarray}
we can build any symmetry for the radial wave function, $(P_xP_y)=(SS)$, $(SA)$, $(AS)$ and $(AA)$,
\begin{eqnarray}
\label{redu}
(SS)&\Rightarrow& R_1^i=\vec{\alpha}_{SS}\cdot\vec{G}_s^i\\ \nonumber
(SA)&\Rightarrow& R_2^i=\vec{\alpha}_{SA}\cdot\vec{G}_s^i\\ \nonumber
(AS)&\Rightarrow& R_3^i=\vec{\alpha}_{AS}\cdot\vec{G}_s^i\\ \nonumber
(AA)&\Rightarrow& R_4^i=\vec{\alpha}_{AA}\cdot\vec{G}_s^i \, ,
\end{eqnarray}
including all possible relative orbital angular momenta
coupled to an S wave. The radial wave function described 
in this section is adequate to describe 
not only bound states, but also it is flexible enough to describe states of 
the continuum within a reasonable accuracy~\cite{Suz98,Vij09,Via09}.

The numerical method described in this section has been successfully tested
in different few-body calculations in comparison with the hyperspherical harmonic 
formalism, see for example Refs.~\cite{Via09,Vin09}, or the 
stochastic variational approach of Ref.~\cite{Suz98} for some of
the results presented in Ref.~\cite{Vij16}.

\section{Results and discussion} 
\begin{figure*}[t]
\resizebox{8.cm}{12.cm}{\includegraphics{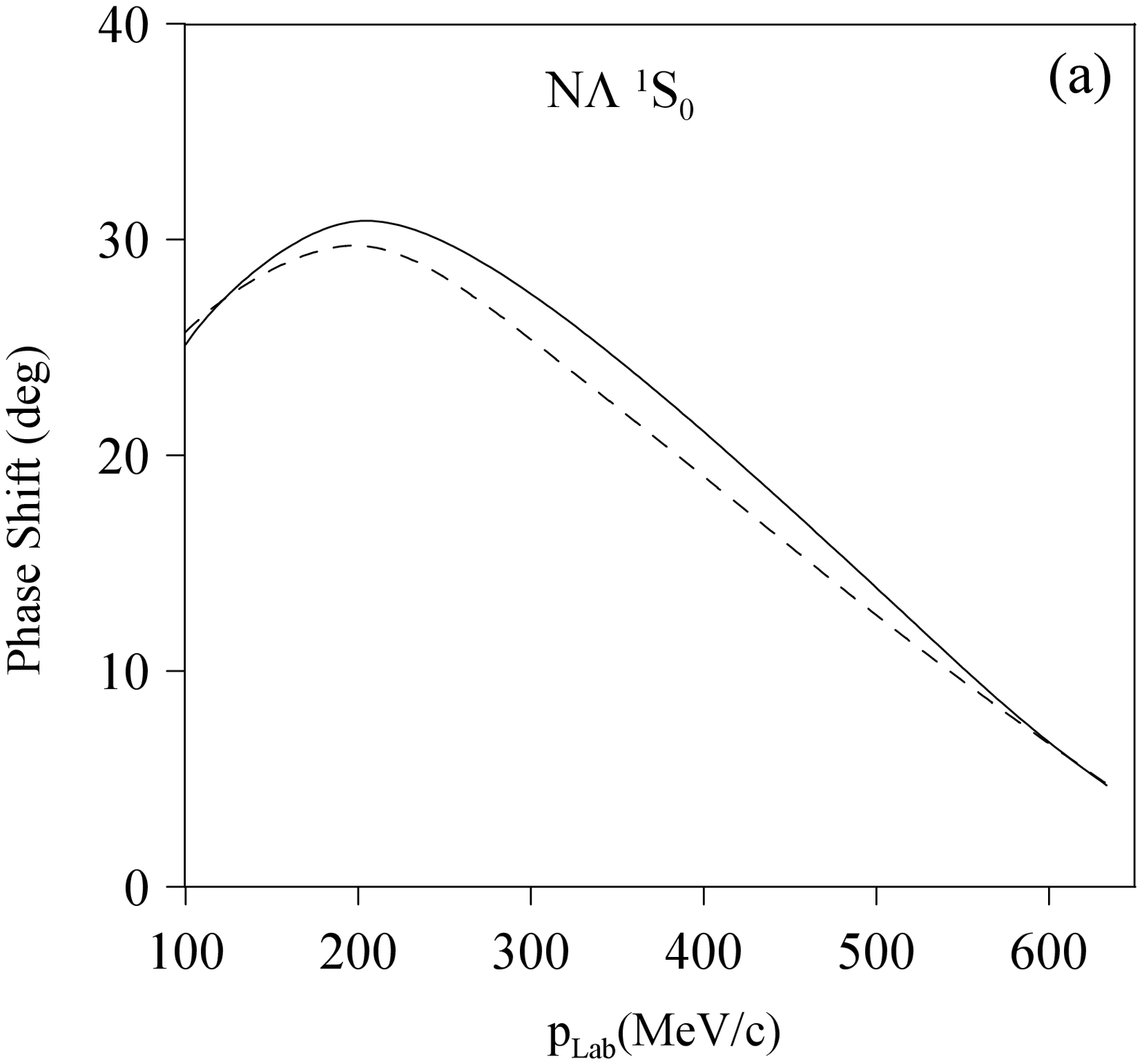}}
\resizebox{8.cm}{12.cm}{\includegraphics{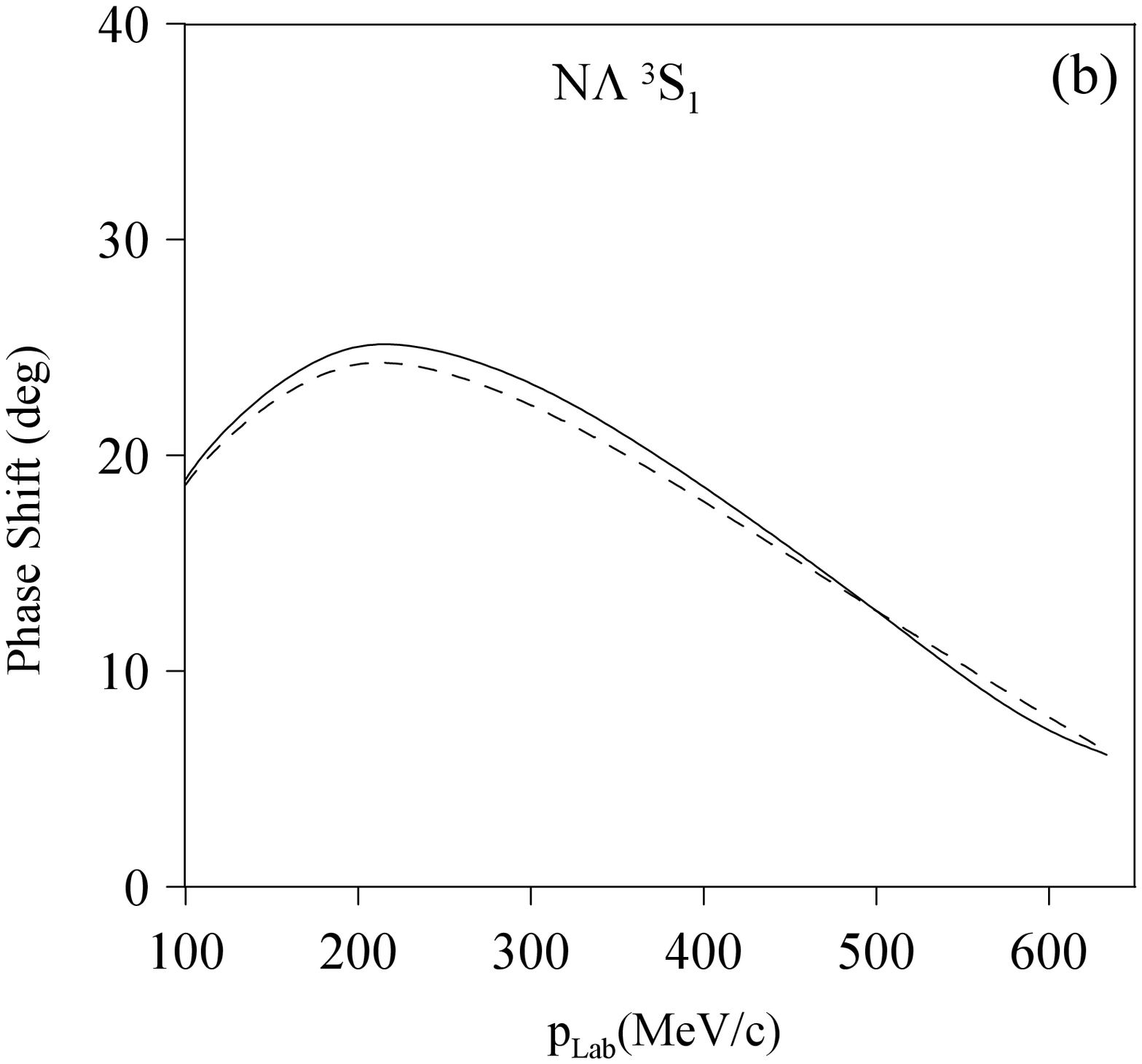}}\vspace*{-5.0cm}
\resizebox{8.cm}{12.cm}{\includegraphics{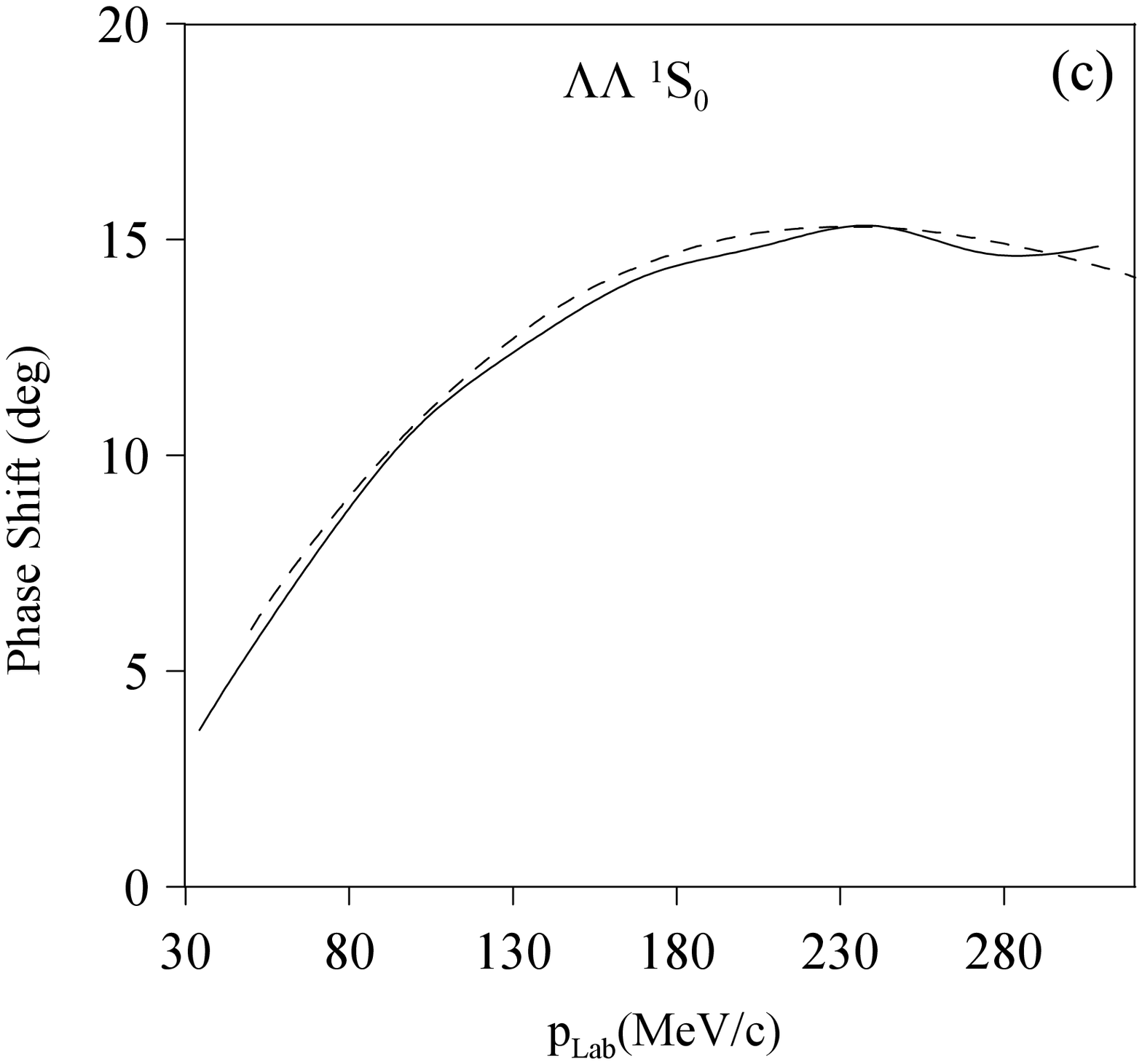}}
\vspace*{-5.0cm}
\caption{(a) $N\Lambda$ $^1S_0$ phase shifts. The solid line stands for the
results of the ESC08c Nijmegen potential and the dashed line for the results
of the two-body potential of Eq.~(\ref{eq21}) with the parameters given in 
Table~\ref{t1}. (b) Same as (a) for the $N\Lambda$ $^3S_1$ phase shifts.
(c) Same as (a) for the $\Lambda\Lambda$ $^1S_0$ phase shifts.}
\label{fig-new}
\end{figure*}

Let us first of all show the reliability of the input potentials.
We compare in Fig.~\ref{fig-new} the $N\Lambda$ and $\Lambda\Lambda$
phase shifts reported by the ESC08c Nijmegen
potential and those obtained by our fits with the two-body potential of 
Eq.~(\ref{eq21}) and the parameters given in Table~\ref{t1}. As can be
seen the agreement is good. Once we have described the phase shifts,
the $N\Lambda$ and $\Lambda\Lambda$ potentials include in an effective 
manner the coupling to other two-body channels as it may be the $N\Sigma$
or $N\Xi$ two-body systems.

We have also tested the two-body interactions in the three-body problem of 
systems made of $N$'s and $\Lambda$'s.
The hypertriton is bound by 144 keV, and the $nn\Lambda$ system is unbound.
The reasonable description on the two- and three-body problem gives confidence
to address the study of the $nn\Lambda\Lambda$ state.

Using the variational method described in the last section, we have evaluated
the binding energy of the $nn\Lambda\Lambda$ system with quantum numbers
$(I)J^P=(1)0^+$. The system is unbound appearing just above threshold and thus
it does not seem to be Borromean, a four-body bound
state without two- or three-body stable subsystems. An unbound result was also
reported in Ref.~\cite{Lek14}, although in this case the authors made use
of repulsive gaussian-type potentials for any of the
two-body subsystems (see the figure on pag. 475) what does not allow
for the existence of any bound state. 
\begin{figure*}[t]
\vspace*{-.5cm}
\resizebox{9.cm}{13.cm}{\includegraphics{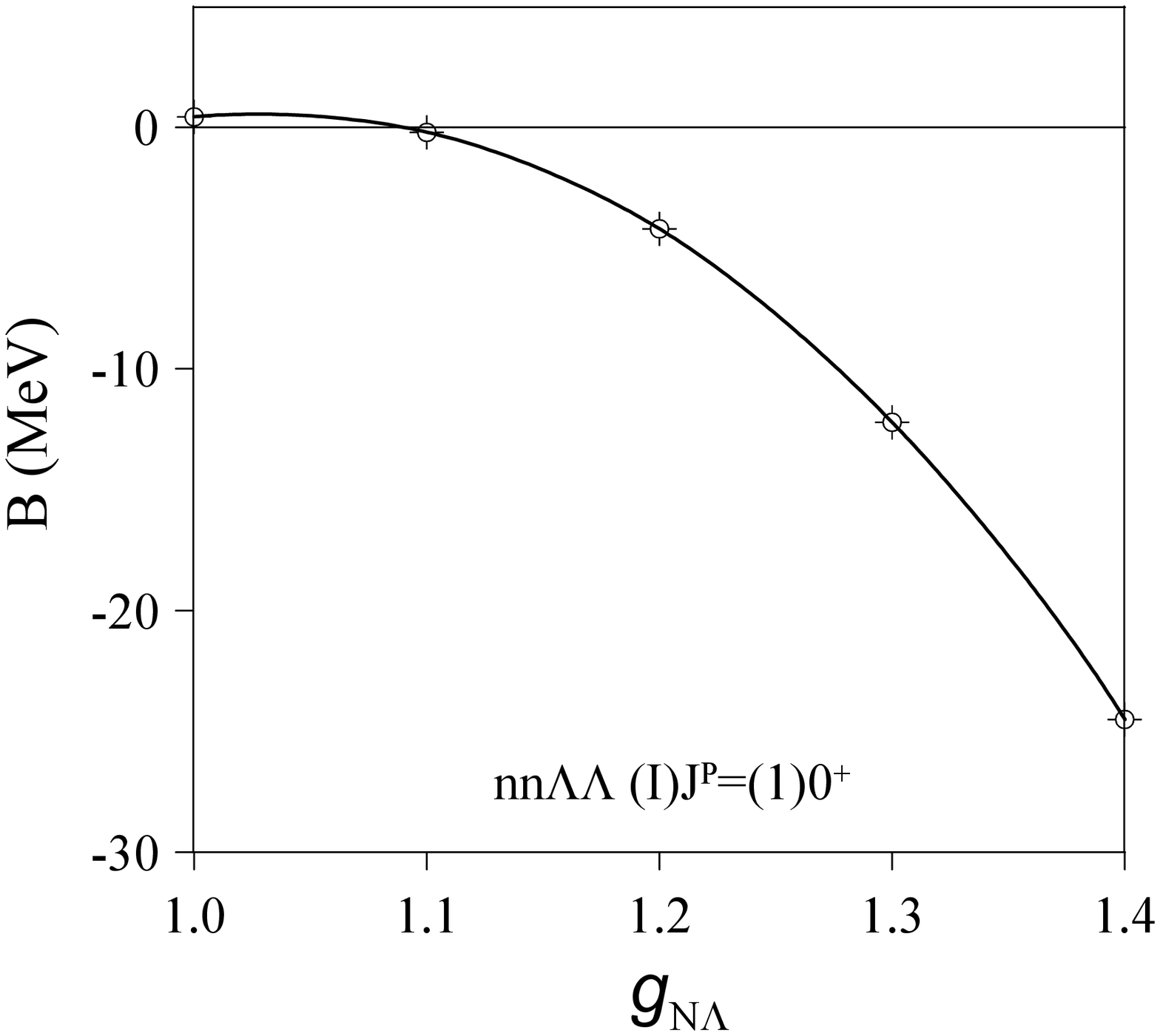}}
\vspace*{-5.0cm}
\caption{Binding energy of the $(I)J^P=(1)0^+$ $nn\Lambda\Lambda$ state as a function of
the multiplicative factor, $g_{N\Lambda}$, in the attractive part of $V^{N\Lambda}(r)$ interaction 
for $g_{NN}=g_{\Lambda\Lambda}=1$.}
\label{fig1}
\end{figure*}

We have studied the dependence of the binding on the strength of
the attractive part of the different two-body interactions entering 
the four-body problem. For this purpose we have used the following
interactions,
\begin{equation}
V^{B_1B_2}(r)=-g_{B_1B_2} \, A\,\frac{e^{-\mu_Ar}}{r}+B\, \frac{e^{-\mu_Br}}{r} \, 
\label{eq21new} 
\end{equation}
with the same parameters given in Table~\ref{t1}.
The system hardly gets bound for a reasonable increase of the strength of the attractive part of
the $\Lambda\Lambda$ interaction, $g_{\Lambda\Lambda}$. 
Although one cannot exclude that the genuine $\Lambda\Lambda$ interaction in 
dilute states as the one studied here could be slightly stronger that the one 
reported in Ref.~\cite{Nag15}, however, one needs $g_{\Lambda\Lambda} \ge 1.8$ to
get a bound state, what would destroy the agreement with the ESC08c Nijmegen 
$\Lambda\Lambda$ phase shifts. Note also that this is also a very sensitive 
parameter for the study of double-$\Lambda$ hypernuclei~\cite{Nem03}. 
Taking a factor $1.2$ in the attractive part of the $^1S_0$ $NN$ interaction, that 
would make the $^1S_0$ $NN$ potential as strong as the $^3S_1$~\cite{Mal69} and 
thus the singlet S wave would develop a dineutron bound state, the four-body system
would start to be bound. 
The situation is slightly different when dealing with the $N\Lambda$ interaction. 
We have used a common factor $g_{N\Lambda}$ for attractive part of the two $N\Lambda$
partial waves, $^1S_0$ and $^3S_1$. We show in Fig.~\ref{fig1} 
the binding energy of the $(I)J^P=(1)0^+$ $nn\Lambda\Lambda$ state as a function of
the multiplicative factor $g_{N\Lambda}$, for $g_{NN}=g_{\Lambda\Lambda}=1$.
As one can see the four-body system develops a bound state for $g_{N\Lambda}=1.1$.

Ref.~\cite{Ric15} studied the 
${}_{\Lambda\Lambda}^{\,\,\,\,4}n$ system based on the fit of
Nijmegen-RIKEN~\cite{Rij10,Rij13} or chiral effective field 
theory~\cite{Pol07} low-energy parameters by means of
a single Yukawa attractive term or a Morse parametrization. 
The method used to solve the four-body problem is similar to the one
we have used in our calculation, thus the results might be directly 
comparable. Our improved description of the two- and three-body subsystems
and the introduction of the repulsive barrier for the $^1S_0$ $NN$ partial wave,
relevant for the study of the triton binding energy (see Table II 
of Ref.~\cite{Mal70}), leads to a four-body state just above threshold,
that cannot get bound by a reliable modification in the
two-body subsystems. As clearly explained in Ref.~\cite{Ric15}, the window of
Borromean binding is more an more reduced for potentials 
with harder inner cores. 

As already discussed in Ref.~\cite{Ric15}, many effects are still to 
be taken into account after arriving
to any definitive conclusion. Among the refinements that would 
eliminate uncertainties, it would be a future challenge to
consider three-body forces that may have an attractive component as 
suggested when studying the triton and $^4$He~\cite{Bev77}.
Although by fitting the $N\Lambda$ phase shifts, the coupling to the 
$N\Sigma$ system has been included in an effective manner,
it would also be interesting to unfold the effective $\Lambda N$ interaction,
separating the contribution from $\Lambda N \leftrightarrow \Sigma N$.
As it has been discussed in the literature ~\cite{Gar14,Hiy14,Gar07,Miy95,Gib77,Gib79} 
the hypertriton does not get bound by considering only $NN\Lambda$ channels, but it is 
necessary to include also $NN\Sigma$ channels.
Similar considerations hold from the $\Lambda\Lambda \leftrightarrow N\Xi$ coupling,
that is expected to play a minor role in this case, because the nucleon
generated in the transition must occupy an excited $p-$shell, the lowest 
$s-$shell being forbidden by the Pauli principle~\cite{Nem03,Gar13}.

\section{ Summary} 
In brief, based on a reasonable approach to the interactions of 
two-body subsystems contributing to the $(I)J^P=(1)0^+$ $nn\Lambda\Lambda$
state, it does not present a bound state. 
We have fitted not only the low-energy parameters of the two-body 
subsystems, but also the phase-shifts. We have considered the repulsive barrier
in the two-body interactions, that it is relevant for a correct 
description of the triton binding energy. We have also studied the 
strange three-body subsystems involved in the problem, the hypertriton
bound by 144 keV, and the $nn\Lambda$ system that it is unbound. 
Thus, the ${}_{\Lambda\Lambda}^{\,\,\,\,4}n$ four-body system does 
not seem to be Borromean.
Finally, although our arguments on the unbound nature of the 
${}_{\Lambda\Lambda}^{\,\,\,\,4}n$ are strong, one
should bear in mind how delicate is the few-body problem in 
the regime of weak binding, as demonstrated in Ref.~\cite{Nem03}.

\section{Acknowledgments} 
This work has been partially funded by COFAA-IPN (M\'exico), 
by Ministerio de Econom\'\i a, Industria y Competitividad 
and EU FEDER under Contracts No. FPA2013-47443, FPA2015-69714-REDT and FPA2016-77177,
by Junta de Castilla y Le\'on under Contract No. SA041U16,
and by Generalitat Valenciana PrometeoII/2014/066.

\end{document}